\documentclass[11pt,a4paper]{article}
\usepackage[T1]{fontenc}
\usepackage[utf8]{inputenc}
\usepackage[english]{babel}
\usepackage{lmodern}
\usepackage[margin=2.5cm]{geometry}
\usepackage{amsmath,amssymb}
\usepackage{graphicx}
\usepackage{booktabs,tabularx,array}
\usepackage{microtype}
\usepackage[section]{placeins}
\usepackage[font=small,labelfont=bf]{caption}
\usepackage[hidelinks]{hyperref}
\hypersetup{
pdftitle={Visibility-Graph-Based Relative Strength Indicator for Financial Market Decision-Making},
pdfauthor={Rafał Rak},
pdfsubject={},
pdfkeywords={},
pdfcreator={},
pdfproducer={}
}
\pdftrailerid{}
\title{Visibility-Graph-Based Relative Strength Indicator for Financial Market Decision-Making}
\author{Rafa{\l} Rak\\[0.4em]
\small Institute of Physics, Faculty of Exact and Technical Sciences,\\
\small University of Rzeszow, Pigonia 1, 35-310 Rzeszow, Poland\\
\small \href{mailto:rrak@ur.edu.pl}{rrak@ur.edu.pl}}
\date{}
\begin{document}
\maketitle
\begin{abstract}
This study introduces the Visibility-Graph-Based Relative Strength Index (VGRSI), a causal indicator that selects and implicitly weights price increments using backward natural visibility. Independent parameters specify the aggregation window and the backward visibility range. Two aggregation variants use the magnitudes and counts of positive and negative contributions to produce values on a 0--100 scale. The VGRSI was tested in an automated walk-forward experiment based on 30-day optimisation windows followed by {7-day walk-forward windows}. This study covered DJI30, EUR/USD, and XAU/USD over 503 trading days in 2024--2025. {The VGRSI strategy produced positive cumulative net profits for all three instruments, whereas the tested classical-RSI strategy and the VGRSI variant without visibility selection produced cumulative losses under the same walk-forward procedure and trading conditions. Because the analysis covered only three instruments and one historical period, the results should be treated as preliminary.
}
\end{abstract}

\section{Introduction}

{Financial price fluctuations exhibit non-Gaussian distributions, temporal dependence, multiscale organisation, and~multifractal properties~\cite{Mandelbrot1989,StanleyMeakin1988,Muzy1994,Oswiecimka2005,Kwapien2005}. Their structure depends not only on the magnitudes of successive changes but also on how these changes are arranged in time. A~technical indicator may therefore lose relevant information when it uses only local price changes and simple~averages.

{Classical technical indicators, including moving averages, MACD, Bollinger Bands, and~the Relative Strength Index (RSI), remain widely used because they are simple, interpretable, and~easy to implement in automated trading systems. Empirical evidence on the performance of technical trading rules and the RSI is inconsistent~\cite{ParkIrwin2007,SullivanTimmermannWhite1999,NeelyWeller2012}. Muruganandan~\cite{Muruganandan2020} found that an RSI-based trading rule failed to generate a positive return for the BSE Sensex even before transaction costs were deducted. Coe and Laosethakul~\cite{CoeLaosethakul2010} likewise found no consistent outperformance of passive investment using the investigated technical rules, including the RSI, in~a large sample of U.S. stocks. In~contrast, Chong~et~al.~\cite{Chong2014} found that selected RSI configurations generated significant abnormal returns in some OECD markets, although~the results were not consistent between markets. Coakley~et~al.~\cite{Coakley2016} reported that RSI-based foreign-exchange rules remained profitable among currencies in the 2006--2015 subperiod after controlling for data-snooping bias. In~a broader study, Rink~\cite{Rink2023} examined 6406 technical trading rules, including 600 RSI variants, and~found that the performance of the recently best-performing rules generally was not maintained outside of the sample and was strongly reduced by moderate transaction costs. The~direct empirical studies of RSI cited above used daily data, whereas the present study used data sampled at one-, five-, and~thirty-minute intervals, that is, at~substantially higher frequencies. These studies indicate that the performance of the classical RSI is unstable and depends on the market, period, parameter configuration, and~transaction costs. Therefore, the~literature does not show that the RSI is always ineffective, but~also does not confirm that its effectiveness is stable.}

{The classical RSI, introduced by Wilder~\cite{Wilder1978}, aggregates upward and downward price changes without explicitly encoding the nonlocal geometric relations between the corresponding observations in the time--price plane.} Multifractal analyses and studies of market persistence, volatility, trading activity, and~cross-correlations show that the organisation of fluctuations can carry information that is not represented by simple local averages~\cite{CalvetFisher2002,BacryDelourMuzy2001,Lux2008,AusloosIvanova2002,TurielPerezVicente2005,Rak2005,Rak2015,Drozdz2000,Gillemot2006,Podobnik2009,GrechMazur2004,BroutyGarcin2024}.

Network representations are widely used to describe complex systems~\cite{barabX}.~Visibility graphs provide a direct geometric transformation of a time series into a network. In~the natural visibility graph introduced by Lacasa~et~al.~\cite{Lacasa2008}, the~observations are represented by nodes and are connected when the line joining them is not obstructed by intermediate observations. The~horizontal visibility graph is a related and geometrically simpler construction~\cite{Luque2009}.~Visibility-based methods have been applied to exchange rates~\cite{Yang2009}, financial-market integration~\cite{Zhuang2014}, the~S\&P 500 index~\cite{Vamvakaris2018}, and~Bitcoin price fluctuations~\cite{Liu2020}. Natural visibility encoding was also used as an input representation for stock trend prediction~\cite{Huang2021}. These studies show that visibility relations can describe the structural properties of financial time series, although~they do not define an RSI-type oscillator based directly on the direction and magnitude of visibility-selected price~increments.

In this study, we introduce the Visibility-Graph-Based Relative Strength Index (VGRSI). The~indicator combines the 0--100 scale of the RSI with a causal backward natural visibility relation. Only observations available at the time of calculation are used. Positive and negative price increments are selected by visibility. The~same increment may contribute more than once when its node is visible from several later nodes in the aggregation window. This multiplicity acts as a geometric weight. We defined the indicator mathematically and tested it in a rolling walk-forward experiment for three financial instruments. The~analysis is exploratory and is not intended to prove universal profitability or superiority to other indicators. {The empirical evaluation also includes comparisons with a classical-RSI trading strategy and a VGRSI variant without visibility selection.}}

\section{Materials and~Methods}

\subsection{Definition of the VGRSI}

{Let $p_t$, $t=0,\ldots,N$, denote a discrete financial price series, where $p_t$ is the selected price, such as the closing price of the observation recorded at time $t$. The~construction uses two independent parameters. The~aggregation window parameter $W_S$ specifies the number of successive reference points over which visibility-selected contributions are collected, whereas the visibility window parameter $W_V$ specifies the maximum backward range within which visibility is~examined.

For a given instant $t$, the~reference indices form the aggregation set
\[
\mathcal{J}_t^{(W_S)}
=
\left\{\max(0,t-W_S+1),\ldots,t\right\}.
\]

For every $j\in\mathcal{J}_t^{(W_S)}$, a~past point $(i,p_i)$ is considered when $\max(1,j-W_V)\leq i<j$. The~restriction $i\geq 1$ ensures that the one-step increment assigned to point $i$ is defined. ~Point $(i,p_i)$ is visible from $(j,p_j)$ if every intermediate point $(k,p_k)$ lies strictly below the line segment joining them:
\[
p_k<p_j+\frac{p_i-p_j}{i-j}(k-j),
\qquad i<k<j.
\]
Accordingly, the~backward visibility set is
\begin{equation*}
\begin{aligned}
\mathcal{V}_j^{(W_V)}=
\Bigl\{i\in[\max(1,j-W_V),j-1]\cap\mathbb{Z}:\;&
p_k<p_j+\frac{p_i-p_j}{i-j}(k-j)\\
&\text{for every }k\in\mathbb{Z}\text{ such that }i<k<j
\Bigr\}.
\end{aligned}
\end{equation*}
All points satisfying this condition within the range $W_V$ are retained; no additional arbitrary subset of visible points is~selected.

The one-step price increment associated with index $i$ is
\[
\Delta p_i=p_i-p_{i-1},\qquad i\geq1.
\]
Visibility determines which indices enter the calculation, whereas $\Delta p_i$ always refers to two consecutive observations, $i-1$ and $i$.

To separate positive and negative contributions, we define two sets of ordered pairs:
\[
\mathcal{P}_t^{+}=
\left\{(j,i):j\in\mathcal{J}_t^{(W_S)},\ i\in\mathcal{V}_j^{(W_V)},\ \Delta p_i>0\right\},
\]
\[
\mathcal{P}_t^{-}=
\left\{(j,i):j\in\mathcal{J}_t^{(W_S)},\ i\in\mathcal{V}_j^{(W_V)},\ \Delta p_i<0\right\}.
\]
The corresponding sums of positive and negative price-change magnitudes are
\[
S^{+}(t)=\sum_{(j,i)\in\mathcal{P}_t^{+}}\Delta p_i,
\qquad
S^{-}(t)=\sum_{(j,i)\in\mathcal{P}_t^{-}}\left|\Delta p_i\right|,
\]
and their numbers are
\[
N^{+}(t)=\left|\mathcal{P}_t^{+}\right|,
\qquad
N^{-}(t)=\left|\mathcal{P}_t^{-}\right|.
\]
Here, $\left|\mathcal{P}_t^{+}\right|$ and $\left|\mathcal{P}_t^{-}\right|$ denote the cardinalities, i.e.,~the numbers of elements, of~the corresponding sets. The~vertical bars do not denote absolute values in these two~expressions.

Because elements of $\mathcal{P}_t^{+}$ and $\mathcal{P}_t^{-}$ are ordered pairs $(j,i)$, a~given increment $\Delta p_i$ may contribute more than once when its node is visible from several different reference nodes $j$. The~number of such visibility relations therefore provides an implicit geometric weight for that~increment.

When both positive and negative contributions are present, two relative-strength coefficients are defined:
\[
r_S(t)=\frac{S^{+}(t)}{S^{-}(t)},
\qquad
r_N(t)=\frac{N^{+}(t)}{N^{-}(t)}.
\]
In the VGRSI, $r_S$ is the closest counterpart of the relative-strength ratio in the classical RSI. However, $S^{+}$ and $S^{-}$ are constructed from visibility-selected contributions and their multiplicities rather than from Wilder-smoothed gains and losses~\cite{Wilder1978}.}
Two aggregation variants are then considered. Variant A0 averages the relative contributions of magnitude and~count:
\[
r_{A0}(t)=\frac{1}{2}\left[r_S(t)+r_N(t)\right].
\]
Variant A1 divides the magnitude ratio by the count ratio:
{\[
r_{A1}(t)=\frac{r_S(t)}{r_N(t)}=\frac{S^{+}(t)/N^{+}(t)}{S^{-}(t)/N^{-}(t)}.
\]
Thus, whenever both signs occur, A1 is the ratio of the mean positive increment magnitude to the mean negative increment magnitude among the visibility-selected~contributions.

The final VGRSI value is defined on a 0--100 scale, including the boundary cases:
\[
\mathrm{VGRSI}_{r_A}(t)=
\begin{cases}
100, & N^{+}(t)>0\ \text{and}\ N^{-}(t)=0,\\[3pt]
0, & N^{+}(t)=0\ \text{and}\ N^{-}(t)>0,\\[3pt]
50, & N^{+}(t)=N^{-}(t)=0,\\[5pt]
100-\dfrac{100}{1+r_A(t)}, & N^{+}(t)>0\ \text{and}\ N^{-}(t)>0,
\end{cases}
\]
where $r_A(t)$ denotes either $r_{A0}(t)$ or $r_{A1}(t)$. A~value of 100 is assigned when all non-zero contributions are positive, a~value of 0 when all are negative, and~the neutral value of 50 when no non-zero contribution occurs.
}

\begin{figure}[!htbp]
\centering
\includegraphics[width=.999\linewidth]{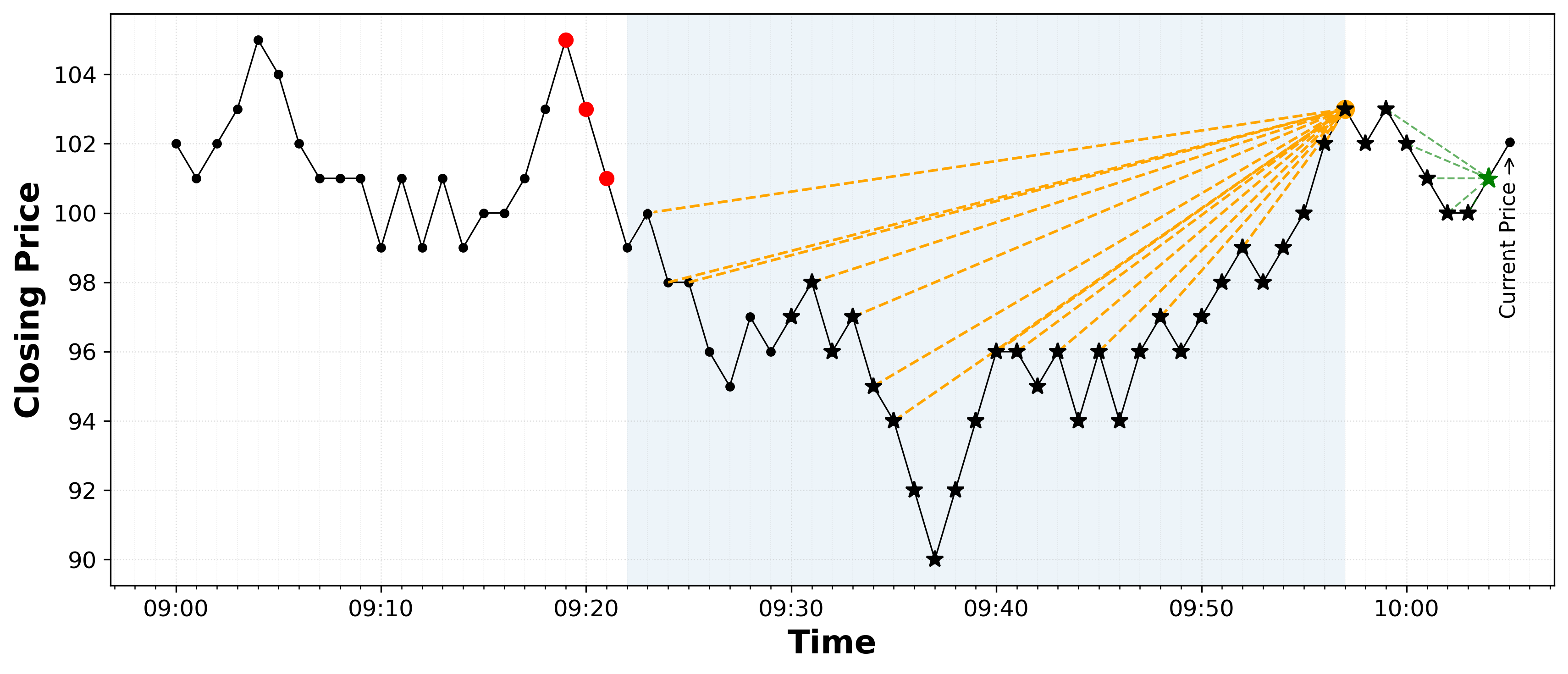}
\caption{Example  of backward visibility for a time series of the closing price of a financial instrument. The~window size $W_S$ = 35 (asterisk symbol) and window visibility $W_V$ = 35 (shaded blue area). The black line and circles show the closing-price series. The orange and green points mark the two reference points for which visibility is illustrated; the matching dashed lines connect them to visible observations. The red circles mark observations visible from the orange point but outside its backward visibility range; these observations are excluded from the calculation for that reference point. The~value of $\mathrm{VGRSI}_{r_{A0}}$ at the current price is 60.1.}
\label{fig1}
\end{figure}

\begin{figure}[!htbp]
\centering
\includegraphics[width=.99\linewidth]{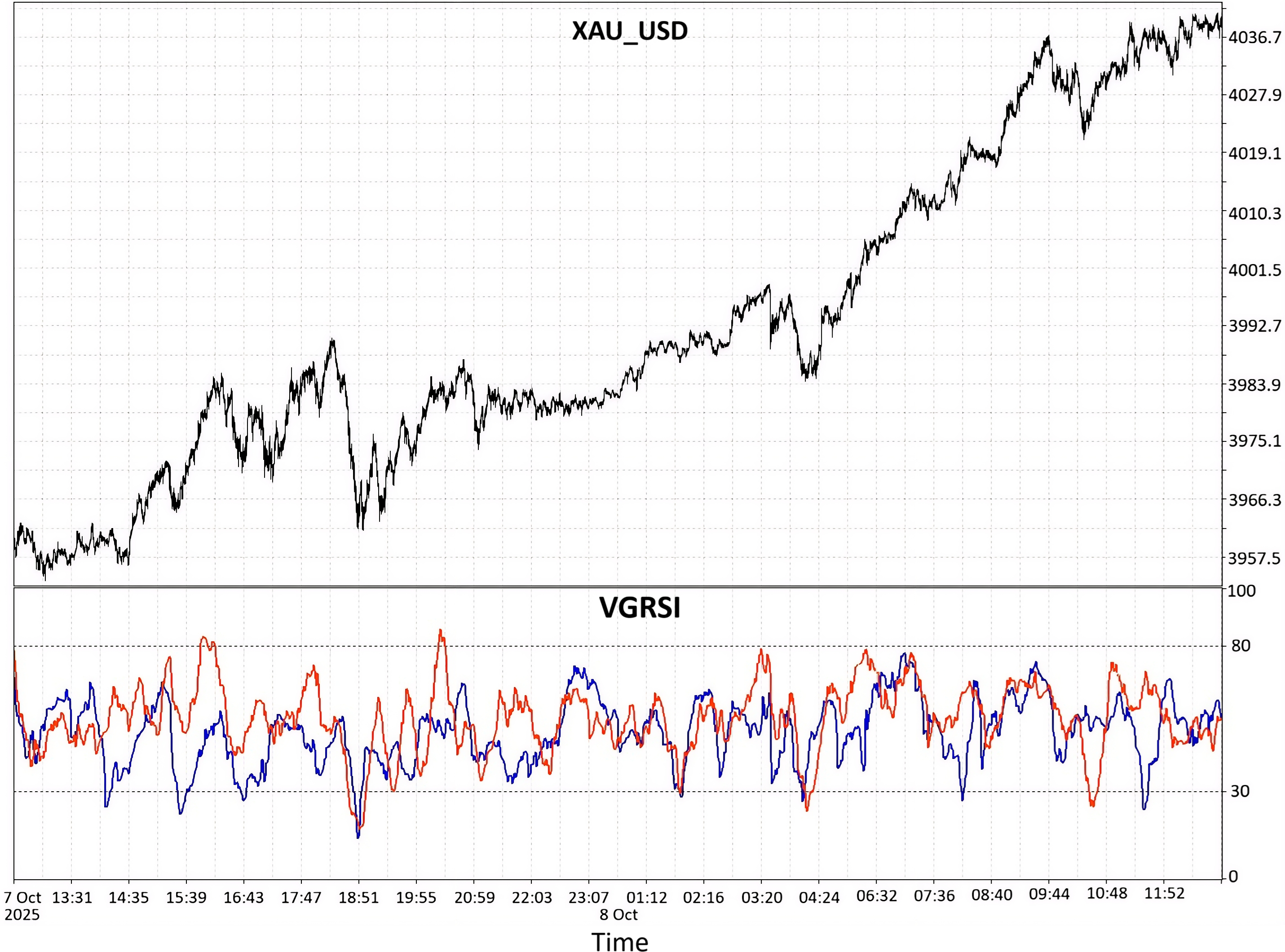}
\caption{The  XAU/USD exchange rates over 7--8 October 2025 (\textbf{top}). The~VGRSI profile for the XAU/USD data (\textbf{bottom}). The~parameters $W_S = 20$, $W_V = 40$, $r_{A0}$ (red line) and $W_S = 15$, $W_V = 100$, $r_{A1}$ (blue line) are used. The black curve in the top panel shows the XAU/USD price. The horizontal dashed lines in the bottom panel mark the reference levels of 30 and 80.
}
\label{fig2}
\end{figure}

{Figure~\ref{fig1} illustrates the backward visibility for $W_S=W_V=35$. The~calculation uses only current and past observations, so the construction is causal. The~time evolution of the VGRSI for XAU/USD is shown in Figure~\ref{fig2}; the two profiles illustrate that A0 and A1 can respond differently to the same price trajectory.
}

\subsection{Interpretation of~the VGRSI}

{{The VGRSI retains the 0--100 scale of the RSI but~selects and weights local price increments using backward visibility.} The parameter $W_V$ controls how far the geometric relation is examined, whereas $W_S$ controls how many successive reference nodes contribute to the final value. The~ordered-pair construction gives greater multiplicity to increments whose nodes remain visible from several subsequent~observations.

Variant A0 combines two forms of directional imbalance: the ratio of total positive to total negative magnitude and the ratio of the corresponding counts. {Because the total magnitude ratio $r_S$ depends on both the numbers and the mean magnitudes of the contributions, A0 should be regarded as a heuristic combination of related quantities rather than an average of independent amplitude and frequency components.} Therefore, it increases when positive contributions dominate in magnitude, number, or~both and decreases under the opposite conditions. Variant A1 has a narrower interpretation. Because~the count ratio cancels algebraically, A1 compares the mean magnitude of positive contributions with the mean magnitude of negative contributions. Therefore, it should not be interpreted as an independent measure of frequency dominance or as direct evidence of a change in the {market regime}}.

\subsection{Experimental~Design}

{{The tests were conducted in MetaTrader 5 using a Trade.MT5 demo account provided by a Forex and CFD broker and connected to the Admiral Group Demo server. An~automated Expert Advisor written in MQL5 was used to conduct the tests, while Python 3 scripts  were used to organise and summarise the rolling results. The~Strategy Tester accessed and processed historical tick data made available through the Admiral Group Demo server directly within MetaTrader 5. No separate data file or tick dataset was manually downloaded or prepared for this study. The~server operated in Eastern European Time (EET, UTC+2 and UTC+3 during daylight saving time). The~broker symbols used in the tests were DJI30, EURUSD, and~XAUUSD, representing an equity index CFD, a~currency pair, and~a gold CFD, respectively. The~Strategy Tester was run in the ``Every tick based on real ticks'' mode. However, trading signals were calculated using closed M1, M5, and~M30 bars. The~tester accounted for the floating bid--ask spreads recorded in the historical data, as~well as the commissions and swap charges specified in the broker's instrument settings. The~results apply to the broker data, symbol specifications, spreads, commissions, leverage, and~execution settings used in the tests.}

{This study covered the two full calendar years of 2024--2025, which corresponds to   503 trading days. Each instrument was independently evaluated. The~leverage was 1:100. At~most two positions could be open simultaneously for a given instrument, and~consecutive entries were separated by at least 30 min. The~reference transaction volume was 1.0 lot for EURUSD, corresponding to a nominal position size of EUR~100,000 and requiring approximately USD~1000 of margin under the adopted leverage. The~volumes used for DJI30 and XAUUSD were adjusted so that the margin required for an individual position was approximately the same as that required for 1.0 lot of EURUSD.}

The evaluation followed a rolling walk-forward design. {For each step, the~genetic algorithm built into MetaTrader 5 typically ran approximately 10,000 parameter configurations within a 30-day optimisation window (rather than all possible parameter combinations). On~the computer used for the tests (a standard desktop PC), the genetic optimisation of a single 30-day window typically took approximately 10--20 min. {The optimisation objective was to maximise net profit within each 30-day optimisation window. When parameter sets produced equal profits, the~set with the smallest drawdown was selected. The~selected parameters were then applied in the following 7-day walk-forward window.
} The optimisation and walk-forward windows were defined in calendar days and were identical for all three instruments. The~first 30-day optimisation window covered 1--30 January 2024, and~the corresponding 7-day walk-forward window covered 31 January--6 February 2024. Then, both windows were shifted forward by seven calendar days. Consequently, successive 7-day walk-forward windows did not overlap. {Optimisation was repeated after each window shift. At~each step, a~set of parameters was selected solely using the preceding 30-day window and kept fixed during the following seven-day test. The~reported cumulative profits comprise only the results of these successive forward tests; optimisation window profits are excluded. The~evaluation therefore concerned the complete procedure of repeated parameter selection and subsequent trading.} {The account balance was reset to USD~10,000 at the beginning of each forward window.} The cumulative profit curves therefore represent the sum of the results obtained for successive 7-day walk-forward windows.}

For every open position, the~symmetric Stop Loss (SL) and Take Profit (TP) distances were fixed at the entry. The~Expert Advisor calculated the median height, expressed in points, of~the most recent $N$ bullish and bearish candles. The~multiplication of this median by the parameter $Z$ determined the SL and TP distances. {In the optimisation, $N$ ranged from 5 to 200 in steps of 30, whereas $Z$ ranged from 5 to 30 in steps of 5.}

{For the VGRSI strategy, the~entry signal was based exclusively on VGRSI.} The indicator was evaluated on three time scales: 1 min (M1), 5 min (M5), and~30 min (M30). {For each timescale and each aggregation variant, $W_S$ and $W_V$ were optimised independently over the range of 10--200 {in steps of 40}. For~both $\operatorname{VGRSI}_{A0}$ and $\operatorname{VGRSI}_{A1}$, the~lower entry threshold was varied from 20 to 35 {in steps of 5}, whereas the upper entry threshold was varied from 70 to 95 {in steps of 5}. A~long (buy) position was opened when both indicator variants crossed their respective lower thresholds from above, that is, moved below these thresholds, on~all three timescales. A~short (sell) position was opened when both variants crossed their respective upper thresholds from below, that is, moved above these thresholds, on~all three timescales. In~both cases, entry also required compliance with the maximum position and minimum time constraints. The~conditions for $\operatorname{VGRSI}_{A0}$ and $\operatorname{VGRSI}_{A1}$ were therefore combined using the logical AND operator.}

{The parameter ranges were informed by prior experience and exploratory experiments covering both 2024 and other years. Their restriction reduced the parameter search space and the computational cost of subsequent optimisation, while retaining relatively broad ranges for instrument-specific parameter selection. The~30-day optimisation and 7-day forward windows were chosen arbitrarily; alternative window lengths may be evaluated in future studies. Since the choice of parameter ranges was not entirely independent of the evaluated period, some risk of strategy-level selection bias remained. The~relatively broad ranges nevertheless provide flexibility in parameter selection, thereby reducing this risk.}

{In the classical RSI comparison, the~two VGRSI parameters, $W_S$ and $W_V$, were replaced by the RSI period, which was optimised independently for each timeframe: M1, M5, and~M30.~The~optimisation range was 10--50 bars with a step of 5. The~instruments, evaluation period, walk-forward procedure, entry threshold ranges, and~procedure for selecting the SL and TP parameters remained unchanged.}

{In the no-visibility variant, all preceding observations within the range $W_V$ were included for each reference point, without~applying the visibility criterion. The~instruments, evaluation period, walk-forward procedure, entry threshold ranges, and~procedure for selecting the SL and TP parameters remained unchanged.}

{The reported statistics were recorded for successive 7-day walk-forward windows. They included trading activity, maximum drawdown, and~the Sharpe ratio, automatically calculated by the MT5 Strategy Tester for each test run. According to the platform documentation, the~Sharpe ratio is calculated as the difference between return and the risk-free rate divided by the standard deviation of return. For~each instrument, the~arithmetic mean was calculated using the Sharpe ratios from 7-day walk-forward windows containing at least one transaction. Windows without transactions were excluded. Such cases occurred only rarely during the evaluation. The~resulting value is therefore the mean of the Sharpe ratios calculated by MT5 for the individual 7-day walk-forward windows.}
}

\section{Results and~Discussion}

\begin{figure}[!htbp]
\centering
\includegraphics[width=\linewidth,height=0.73\textheight,keepaspectratio]{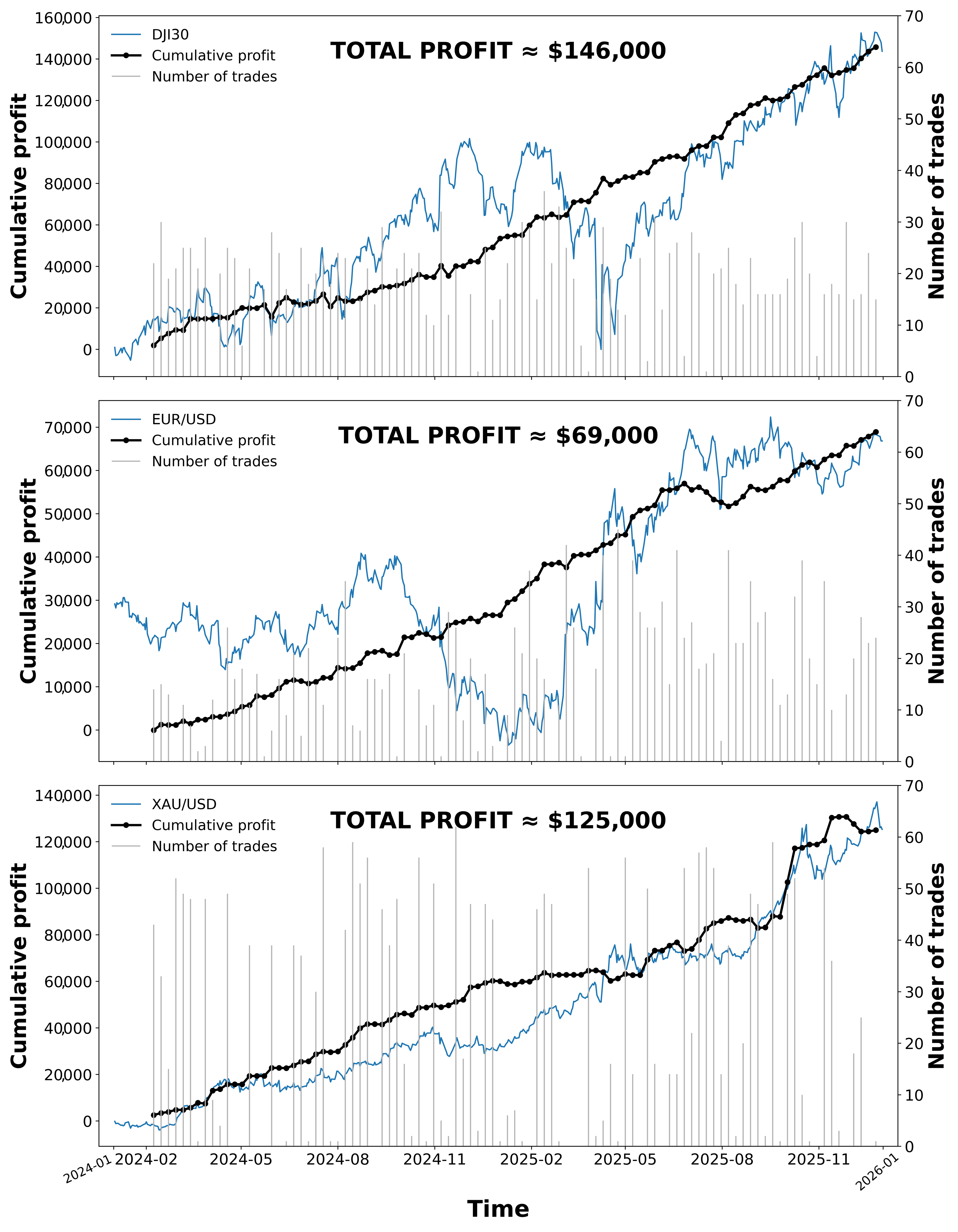}
\caption{Cumulative  profit from successive 7-day walk-forward windows for VGRSI-based strategy applied to broker instruments DJI30 (an equity-index CFD), EURUSD (a currency pair), and~XAUUSD (a gold CFD) from 1 January 2024 to 31 December 2025 (black curves). The~rescaled price series of the analysed instruments are shown in blue, while the grey vertical bars indicate the number of trades executed in each 7-day walk-forward window.}
\label{fig3b}
\end{figure}

{Figure~\ref{fig3b} presents cumulative profit from the successive 7-day walk-forward windows, together with the rescaled instrument prices and the number of trades in each window. The~cumulative result was positive for each instrument independently tested. The~curves also contain flat intervals and local decreases, so the result was not positive in every {7-day walk-forward window}.
}

{{The profits summed across the successive 7-day walk-forward windows were USD~146,000 for DJI30, USD~69,000 for EUR/USD, and~USD~125,000 for XAU/USD (Table~\ref{tab1}). {For the fixed reference capital $C_0=\mathrm{USD}\,10{,}000$, normalised cumulative profit was defined as $\mathrm{NCP}=(\sum_{k=1}^{K}P_k/C_0)\times100\%$, where $P_k$ denotes net profit in the $k$-th forward window, and $K$ is the number of evaluated windows. This measure was approximately 1460\% for DJI30, 690\% for EUR/USD, and~1250\% for XAU/USD. It expresses cumulative profit relative to the fixed test capital and does not assume compounding across windows.}\par}
}

\begin{table}[!htbp]
\centering
\caption{{Summary  of trading activity and results for VGRSI-based strategy. The~numbers of all, long, and~short trades, the~Sharpe ratio calculated by MT5, and~the maximum drawdown were evaluated within fixed 7-day walk-forward windows and then summarised across all windows. The~mean Sharpe ratio was calculated using windows containing at least one transaction; windows without transactions were excluded. Total trades, trades per day, and~the sum of window profits refer to the complete 2024--2025 evaluation period for each independently tested instrument.}}
\label{tab1}

\footnotesize
\renewcommand{\arraystretch}{1.2}
\begin{tabularx}{\linewidth}{l
>{\centering\arraybackslash}X
>{\centering\arraybackslash}X
>{\centering\arraybackslash}X
>{\centering\arraybackslash}X
>{\centering\arraybackslash}X}
\toprule
\textbf{Asset }
& \multicolumn{2}{c}{\textbf{All Trades (Min/Max/Mean)}}
& \multicolumn{2}{c}{\textbf{Long Trades (Min/Max/Mean)}}
& \textbf{Short Trades (Min/\allowbreak Max/\allowbreak Mean)} \\
\midrule
DJI30
& \multicolumn{2}{c}{0/37/18}
&\multicolumn{2}{c}{ 0/32/10}
& 0/29/8 \\
EUR/USD
& \multicolumn{2}{c}{0/45/16}
&\multicolumn{2}{c}{ 0/40/7}
& 0/39/9 \\
XAU/USD
& \multicolumn{2}{c}{0/62/24}
&\multicolumn{2}{c}{ 0/55/18}
& 0/45/6 \\
\midrule
\textbf{Asset}
& {\textbf{Sharpe Ratio (Mean)}}
& {\textbf{Window Max Drawdown \% (Mean)}}
& \textbf{Total Trades}
& \textbf{Trades/Day}
& {\textbf{Sum of Window Profits (USD)}} \\
\midrule
DJI30
& 3.6
& 18
& 1842
& 3.5
& 146,000 \\
EUR/USD
& 2.55
& 12
& 1677
& 3.3
& 69,000 \\
XAU/USD
& 3.20
& 10
& 2418
& 4.8
& 125,000 \\
\bottomrule
\end{tabularx}
\end{table}

DJI30 generated the highest cumulative profit but~also the largest mean maximum drawdown within the seven-day windows. XAU/USD had the second-highest cumulative profit with the lowest mean window-level maximum drawdown among the three instruments.

Trading activity ranged from 3.3 to 4.8 trades per day. Both long and short positions occurred for every instrument. EUR/USD showed a slightly greater mean number of short than long trades, DJI30 exhibited a comparatively balanced distribution, and~XAU/USD showed a clear predominance of long positions. These differences indicate that the selected walk-forward configurations did not produce the same directional exposure for all three~markets.

The cumulative profit curves did not simply reproduce the rescaled price trajectories. For~example, the~cumulative result of EUR/USD increased during parts of the sample during which the instrument price declined. Similar departures between the price and cumulative result curves are visible for DJI30. This observation suggests that the evaluated strategy was not equivalent to passive long-only exposure. It does not, by~itself, establish that the VGRSI is independent of conventional momentum or trend-following~effects.

{The arithmetic means of the Sharpe ratios calculated by MT5 for 7-day walk-forward windows containing at least one transaction ranged from 2.55 to 3.60. {The mean maximum-drawdown percentages recorded within the individual seven-day windows were 18\% for DJI30, 12\% for EUR/USD, and~10\% for XAU/USD. These are averages of window-level statistics rather than maximum drawdowns over the entire evaluation period.
}}

{The account balance was reset to USD~10,000 at the beginning of each forward window. The~annual-scale Sharpe ratio was calculated from the seven-day window returns as follows: $r_k=P_k/C_0$ and $SR_{\mathrm{ann}}=\sqrt{52}\,\bar r/s_r$, where $P_k$ denotes the net profit in the $k$-th 7-day forward window, $C_0=\mathrm{USD}\,10{,}000$, and~$\bar r$ and $s_r$ are the mean and sample standard deviation of the window returns. The~factor $\sqrt{52}$, with~52 denoting the number of weekly periods conventionally assumed per year, provides the conventional conversion of the weekly measure to an annual scale. A~zero risk-free rate was used, and~zero-profit windows were retained in the sequence. The~resulting values were approximately 4.3 for DJI30, 3.71 for EUR/USD, and~3.2 for XAU/USD. The~annual-scale Sharpe measure was highest for DJI30, followed by EUR/USD and XAU/USD, indicating the highest mean window return relative to its variability for DJI30.}

{Figure~\ref{fig4rsi} presents the cumulative profits obtained using the classical RSI strategy. For~the tested variant, these amounted to $-\mathrm{USD}\,1844$ for DJI30, $-\mathrm{USD}\,50{,}921$ for EUR/USD, and~$-\mathrm{USD}\,32{,}904$ for XAU/USD. The~tested RSI strategy therefore ended the evaluation period with a cumulative loss for all three instruments, while the VGRSI strategy produced positive cumulative profits under the adopted test conditions.}

{The VGRSI variant without visibility selection yielded cumulative profits of   \mbox{$-33{,}848$ USD} for DJI30, $-30{,}190$ USD for EUR/USD, and~$-27{,}769$ USD for XAU/USD. This variant therefore produced losses for all three instruments, whereas the VGRSI with visibility selection produced positive cumulative profits. Under~the tested conditions, retaining visibility selection improved the cumulative result for each instrument. {{The cumulative net profits of the three strategies are summarised in Table~\ref{tab2}.}}}

{\begin{table}[!htbp]
\centering
\caption{{Comparison of cumulative net profit (USD) for the three strategies over the 2024--2025 evaluation period. Each value is the sum of net profits from successive seven-day forward windows. The~instruments were evaluated independently.}}
\label{tab2}
\renewcommand{\arraystretch}{1.2}
\begin{tabularx}{\linewidth}{l
>{\centering\arraybackslash}X
>{\centering\arraybackslash}X
>{\centering\arraybackslash}X}
\toprule
\textbf{Asset} & \textbf{VGRSI} & \textbf{Classical RSI} & {\textbf{VGRSI Without}   \textbf{Visibility Selection}} \\
\midrule
DJI30 & 146,000 & $-$1844 & $-$33,848 \\
EUR/USD & 69,000 & $-$50,921 & $-$30,190 \\
XAU/USD & 125,000 & $-$32,904 & $-$27,769 \\
\bottomrule
\end{tabularx}
\end{table}
}

{These results concern the complete VGRSI-based trading strategy and do not establish whether A1 detects breakouts or changes in the market regime}.

{The losses obtained for the classical RSI variant tested should be interpreted in the context of mixed previous evidence. Muruganandan~\cite{Muruganandan2020} found that the examined RSI rule failed to generate a positive return for the BSE Sensex even before transaction costs. In~contrast, Chong~et~al.~\cite{Chong2014} identified profitable RSI configurations in selected markets, including the RSI(14,\ 30/70) rule for the Dow Jones Industrial Index. Coakley~et~al.~\cite{Coakley2016} also reported profitable RSI-based foreign exchange rules after controlling for data-snooping bias. Therefore, the~present findings do not support the general conclusion that the RSI is ineffective. These studies used daily data, whereas our strategy uses M1, M5, and~M30 signals. Differences in periods, instruments, entry and exit rules, transaction costs, and~parameter selection procedures may have contributed to the different outcomes; their individual effects were not tested in the present~study.

\begin{figure}[!htbp]
\centering
\includegraphics[width=\linewidth,height=0.73\textheight,keepaspectratio]{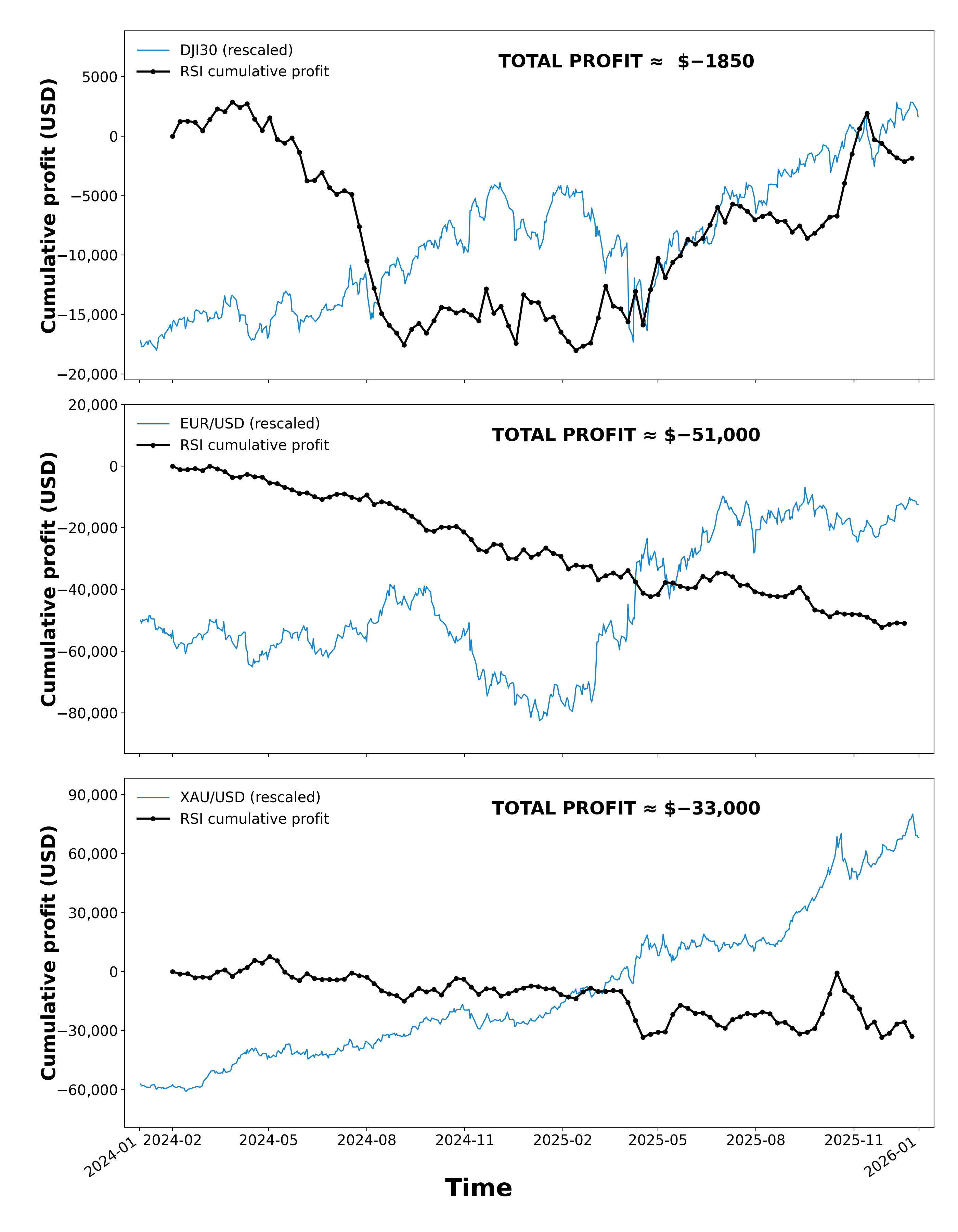}
\caption{{Cumulative   profits from successive 7-day walk-forward windows for classical RSI strategy (black curves). Blue curves show the rescaled daily closing prices of DJI30, EUR/USD, and~XAU/USD. The~walk-forward procedure, trading settings, and~ranges of parameters shared by the two strategies were retained from the VGRSI experiment shown in Figure~\ref{fig3b}.}}
\label{fig4rsi}
\end{figure}

The VGRSI is also related to earlier financial applications of visibility graphs. Vamvakaris~et~al.~\cite{Vamvakaris2018} used horizontal visibility graphs to investigate changes in S\&P 500 during crises, while Huang~et~al.~\cite{Huang2021} applied natural visibility encoding to predict {stock trends}. A~common element is the use of temporal and geometric structure in the series. In~the VGRSI, backward visibility directly determines the selection and multiplicity of local increments contributing to a relative-strength oscillator. The~present results concern the profits of a complete walk-forward strategy and are therefore not directly comparable to network structure measures or trend-prediction accuracy. {The comparison with the no-visibility variant assesses the contribution of visibility selection within the tested \mbox{trading procedure.}}
}

{In the three independent simulations, the~sum of profits from successive 7-day walk-forward windows was USD~340,000.} In each step, the parameters selected in a 30-day window were applied to the next 7 days. This separated optimisation from the immediately following evaluation period. However, it did not remove all the risks of selection bias. The~result may depend on the parameter ranges, the~optimisation criterion, and~the selected \mbox{30/7-day}~schedule.

The experiment was limited in scope. It included three instruments, one two-year period, and~one automated trading design. {The additional classical RSI comparison was limited to the variant and settings tested.} {{These comparisons do not establish superiority for other settings and indicators.}}

The economic interpretation is also conditional on the implementation. The~three instruments were tested independently, with~separate initial balances; therefore, the~results do not represent a combined tradable portfolio. {{The present study reports normalised cumulative profit relative to the fixed reference capital, an~annual-scale Sharpe measure calculated from the sequence of window returns, and~window-level drawdown statistics. The~normalised cumulative profit summarises the aggregate outcome over the evaluation period without assuming compounding across account resets. The~retained simulation outputs support these measures but do not provide a sufficiently complete equity history to reconstruct full-period maximum equity drawdown reliably.}} {The backtests used broker-supplied historical ticks and instrument-specific trading costs configured in MT5.}

\section{Conclusions}

{This study introduced the VGRSI, a~causal relative-strength indicator that uses natural backward visibility to select and implicitly weight local price increments. Two variants were defined. A0 combines the imbalance in the magnitudes and numbers of positive and negative contributions, whereas A1 compares their mean magnitudes. The VGRSI was implemented in MT5, a~widely available trading platform, and~tested in a rolling walk-forward procedure for DJI30, EUR/USD, and~XAU/USD in 2024--2025. {Under the adopted trading conditions, cumulative net profits were approximately USD~146,000 for DJI30, USD~69,000 for EUR/USD, and~USD~125,000 for XAU/USD. Both the classical RSI strategy and the VGRSI variant without visibility selection produced cumulative losses for all three instruments. {These comparisons indicate that the use of the proposed visibility-graph-based indicator improved cumulative net profits within the adopted \mbox{walk-forward procedure.}}}

{The results were obtained in successive seven-day forward windows, using parameters selected in the preceding 30-day optimisation window. {The interpretation of these results should account for possible selection bias associated with the choice of parameter ranges and the incomplete assessment of risk over the full evaluation period.} {Given the limited scope of this study, the~results are preliminary and do not establish the effectiveness of the indicator for other assets.}} Further research should include more instruments, different optimisation and walk-forward window lengths, full-period risk measures, and~comparisons with other commonly used indicators.
}

\section*{Author Contributions}
Conceptualisation, R.R.; methodology, R.R.; software, R.R.; validation, R.R.; formal analysis, R.R.; investigation, R.R.; resources, R.R.; data curation, R.R.; writing---original draft preparation, R.R.; writing---review and editing, R.R.; visualisation, R.R. The~author has read and agreed to the published version of this~manuscript.

\section*{Funding}
This  research received no external~funding.

\section*{Institutional Review Board Statement}
Not applicable.

\section*{Informed Consent Statement}
Not applicable.

\section*{Data Availability Statement}
{The historical tick data used in this study were accessed and processed directly within MetaTrader 5 through the Admiral Group Demo server. No separate data file or tick dataset was manually downloaded or prepared for this study. Access to the historical data is subject to their availability on the broker's server and to the broker's access conditions. {Selected portions of the strategy code will be made available by the author to readers upon individual request.}}

\section*{Conflicts of Interest}
The author declares no conflicts of~interest.

\end{document}